\documentclass[]{spie}  
\usepackage{array} 
\usepackage{subcaption}
 
\usepackage{amsmath,amsfonts,amssymb}
\usepackage{graphicx}
\usepackage[colorlinks=true, allcolors=blue]{hyperref}
\usepackage{amsmath}
\usepackage{textcomp} 

\title{Frequency-domain Event-based Imaging for Selective Surveillance}

\author{Megan Birch\textsuperscript{1}}
\author{James Rick}
\author{Adrish Kar}
\author{Jason Zutty}
\author{Joseph L. Greene\textsuperscript{2}}
\affil{Georgia Tech Research Institute, 250 14th Street, NW, Atlanta, GA 30318, USA}

\authorinfo{Direct correspondence to: \\1 E-mail: megan.birch@gtri.gatech.edu\\ 2 E-mail: joseph.greene@gtri.gatech.edu}

\begin{document} 
\maketitle

\begin{abstract}
Event-based cameras (EBCs) are emerging as an attractive sensing modality for surveillance applications by encoding pixel-level radiance changes with microsecond resolution and high dynamic range, enabling motion extraction while suppressing static background effects. However, their asynchronous, sparse output requires algorithms that can identify targets directly in event space without processing full image frames. To address this need,  we introduce Frequency Rate Information for Event-Space (FRIES), a neuromorphic processing framework with potential to discriminate and monitor man-made objects by detecting periodicity in pixel-level events, such as those caused by rotor rotation and mechanical vibrations. FRIES begins with a time gating operation that functions to suppress background noise, then aggregates events into a pixel-wise activity (e.g., density) map and clusters pixels into regions of interest (ROIs). A localized spectral analysis is then applied to each ROI to extract a dominant frequency that can be used to discriminate structured object signature detections from unstructured background and noise detections. Discriminated targets are subsequently visualized using a Resonant Time Surface (RTS), a frequency-selective visualization method that weights incoming events based on their phase coherence with extracted frequencies to reward in-sync content while suppressing out-of-sync clutter and noise. FRIES and the RTS were demonstrated with a controlled indoor experiment by recovering the rotational frequency of a mechanical chopper and drone rotors against a moving background. FRIES and the RTS were then tested on outdoor environment data, where a hovering drone was detected against a cluttered treeline. These preliminary results establish frequency-domain event processing as a promising front-end to enable selective surveillance in neuromorphic pipelines, and a promising compliment to systems with conventional framing cameras due to their significantly higher temporal resolution that enables discrimination through spectral analysis.
\end{abstract}


\section{Introduction}
\label{sec:intro}  

Existing surveillance systems exhibit reduced performance when interrogating targets under rapid motion, low-contrast scenes, complex backgrounds, or degraded conditions (e.g., fog, optical turbulence), creating operational gaps in automatic target recognition (ATR) pipelines for real-world deployment \cite{munir_situational_2022}. Further, systems often struggle with discriminating features of confusers for small UAS platforms, such as birds, degrading accuracy. Event-based cameras (EBCs) offer an additional solution to mitigate these challenges by providing a contrast-based encoding scheme that reports pixel-level changes in scene radiance (referred to as events) with microsecond-level resolution and high dynamic range \cite{gallego_event-based_2022}. Under this paradigm, EBCs encode scene changes (e.g., motion, intensity fluctuation) while suppressing static background \cite{gallego_event-based_2022}, reducing motion blur \cite{jiang_learning_2020}, and maintaining performance under low-light \cite{maqueda_event-based_2018} or optically turbulent conditions \cite{boehrer_turbulence_2021}. These properties have enabled EBC-based monitoring in biological imaging \cite{guo_eventlfm_2024, li_e-gaze_2024}, low-light situational awareness \cite{liu_seeing_2024, cohen_event-based_2019}, and object detection \cite{chen_neuroaed_2020}, with particular focus on small or rapidly changing targets such as drones \cite{magrini_drone_2025} or detonation characterization \cite{parab_low_2025}.

However, EBCs exhibit practical challenges when applied to surveillance tasks. First, the asynchronous event streams remain incompatible with traditional image processing pipelines, which assume a fixed frame structure, leading many existing methods to reconstruct intensity-like frames or event histograms before analysis \cite{rebecq_events--video_2019, lakshmi_neuromorphic_2019}, reducing the effectiveness of the contrast-driven event representation. Second, methods that operate natively in event space, such as time surfaces \cite{sironi_hats_2018} and neural networks \cite{cordone_object_2022}, rely on spatial or temporal cues to detect objects, which assumes clustered events must arise from similar structures. In practice, objects, such as those arising from man-made origins, may exhibit periodic structures from mechanical oscillation (e.g., rotating rotors, vibrating motors) \cite{tinney_multirotor_2018}, yielding a spectral signature to distinguish from unstructured background motion or noise. Third, detection performance across EBCs remains sensitive to a user-defined contrast threshold whose optimal value is scene-dependent as well as object-dependent and requires precise knowledge of event camera noise models \cite{greene_pytorch-enabled_2025} that remain difficult to calibrate \cite{paredes-valles_back_2021, muglikar_how_2021}. These challenges motivate the need for algorithms to leverage the rich temporal structure of the event stream beyond spatial statistics, and may operate on unknown scene conditions without extensive sensor calibration.

\begin{figure}[ht]
    \centering
    \includegraphics[width=\linewidth]{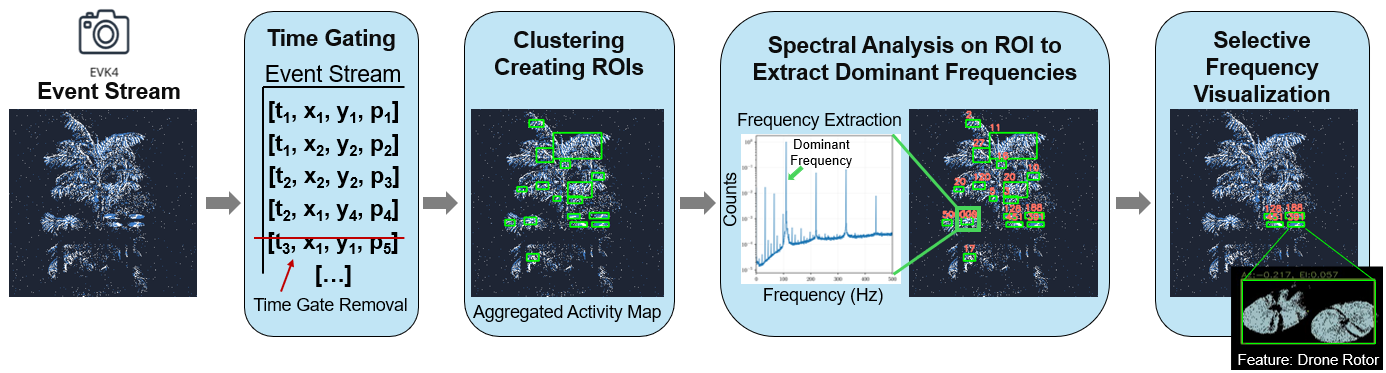}
    \caption{Overview flow chart of the FRIES algorithm. Events are temporally filtered, aggregated into a pixel-wise activity map, and thresholded to suppress background activity. The remaining high-activity pixels are spatially clustered into ROIs, from which dominant frequencies are extracted. These ROI frequency values are then used as a spectral filter to isolate and characterize target objects, such as drone rotors shown in the last step.}
    \label{fig:FlowChart3}
\end{figure}

To address this gap, we introduce \textbf{F}requency \textbf{R}ate \textbf{I}nformation for \textbf{E}vent-\textbf{S}pace (FRIES), a neuromorphic processing framework that leverages the high temporal resolution of the event stream to discriminate man-made targets through their spectral signatures. 
As shown in Figure \ref{fig:FlowChart3}, FRIES operates in four stages. First, a temporal gating operation filters the raw event stream to suppress sensor and background noise while preserving the frequency band of interest, where the selected band is dependent on the target and environment. Second, filtered events are aggregated into a pixel-wise activity map and spatially clustered into Region of Interests (ROIs) based on event density. To reduce sensitivity to sensor calibration and scene-specific priors, activity thresholds are set adaptively relative to the median background rate rather than fixed contrast parameters, allowing FRIES to operate robustly across different sensors and environments without manual retuning. Third, a Fourier transform is applied per pixel across each ROI to extract frequency content across the region. Next, each ROI is assigned a single dominant frequency value determined by the most common frequency peak across all pixels. Finally, the Resonant Time Surface (RTS), a frequency-selective event weighting and visualization tool, weights incoming events in real time based on their phase coherence with the extracted target frequency to selectively enhance in-sync events while suppressing out-of-sync clutter to provide selective surveillance to inform downstream ATR pipelines. To demonstrate FRIES, we collect and process data using a Prophesee EVK4 event-based camera across controlled indoor and uncontrolled outdoor experiments, demonstrating frequency extraction and selective target isolation in cluttered scenes.

\begin{sloppypar}
\end{sloppypar}

\section{Methods}

This section describes the design of the FRIES algorithm. Section \ref{sec:temporal_filter} describes the temporal gating operation, Section \ref{sec:detection} describes the spatial clustering, Section \ref{sec:frequency_extraction} describes the spectral analysis, and Section \ref{sec:rts_vizualization} describes the visualization process through RTS processing.

\subsection{Time Gating}
\label{sec:temporal_filter}
Before filtering, the event stream is separated based on polarity to prevent positive and negative events from canceling when summed, which could otherwise suppress activity at an oscillating edge. This separation preserves the total number of radiance-change events and ensures periodic on–off transitions remain visible in subsequent steps. To allow frequency content to emerge over a sufficient observation window, events are batched into fixed intervals of 10ms. Within each batch, a time gate is applied to each pixel against the most recent timestamp from the previous gated batch in that pixel, i.e. for each pixel, the interval between each event and the previous per pixel timestamp value is computed, and events are only retained if their interval falls within some minimum time $t_{min}$ and maximum time $t_{max}$. Here, $t_{min}=3ms$ and $t_{max}=6ms$ are selected to suppress high-frequency EBC noise artifacts while preserving the frequency content of drone rotors, observed to range from approximately 50-500Hz in the experiments described in Section \ref{sec:results}. 

\subsection{Detection/Cluster}
\label{sec:detection}
Following time gating, the retained events from the batch at each pixel coordinate are summed independently for each polarity channel, and their counted events are combined to form a single per-pixel activity score. These scores are arranged into a 2D activity map that reflects the spatial distribution of event density after the temporal gating operation. Because full spectral analysis through a Fourier transform is computationally expensive relative to a counting operation, the activity map serves as a lightweight pre-screening step such that only pixels whose activity score exceeds a threshold are forwarded for spectral analysis. The threshold is set to the median activity score across the map, providing an adaptive estimate of the background event rate that requires no scene-specific calibration. Using the median value incorporates two assumptions: first, the targets are sparse within the scene such that the median value is reflective of a background level and second, the objects of interest produce events at a higher event rate than the background due to their structured oscillations. The thresholded activity map is then spatially downsampled by 2x in each dimension to reduce computational load before clustering while minimally sacrificing resolution \cite{9400234}. Density-Based Spatial Clustering of Applications with Noise (DBSCAN) is applied to the downsampled thresholded map to group spatially contiguous high-activity pixels into candidate ROIs \cite{khan_dbscan_2014}. DBSCAN is selected for its ability to recover arbitrarily shaped clusters without requiring a predetermined number of clusters, which is well-suited to the irregular spatial footprints of rotating mechanical objects.

\subsection{Spectral Analysis}
\label{sec:frequency_extraction}
For each detection candidate or ROI, events are retrieved from the unfiltered stream and arranged into a per-pixel time series over a uniform time grid, with unoccupied time points being assigned a value of zero to ensure a consistent sampling support for spectral analysis. To account for the effect of the refractory period (e.g., pixel downtime after an event occurs) on the spectral analysis, events need to be convolved with a ramp function with a temporal width of 6.1 microseconds to match the measured refractory period of a Prophesee EVK4 \cite{chassagnol_pixel_2025}. However, we recognize this step serves as an approximation that remains accurate only if the frequency of filtered events remains lower than the refractory period and will otherwise require full analysis from the Laplace transform to resolve \cite{robinson_renewal_1997}. A discrete Fourier transform is applied to each per-pixel time series and converted to a Power Spectral Density (PSD) map to identify the frequency carrying the most power. The PSD map provides a quantitative measure of how event energy is distributed over frequency, allowing stable, dominant oscillatory components to be distinguished from noise and transient fluctuations. The peak-power frequency is extracted per pixel, and each ROI is assigned the most common frequency across its constituent pixels. In the event of a tie, the frequency with the highest aggregate PSD value is selected. This assignment ensures each detection candidate's ROI is labeled by the frequency that most consistently drives its activity rather than by isolated or transient spectral peaks. To promote temporal continuity across successive batches, per-pixel activity scores are carried forward and exponentially decayed over time, smoothing the activity map against transient noise and reinforcing stable, persistent signal sources. The decay rate is treated as a tunable parameter governing the trade-off between responsiveness to new activity and robustness to intermittent noise.

\subsection{RTS Visualization}
\label{sec:rts_vizualization}
RTS is a method for weighting and visualizing incoming events based on a frequency-selective filtering method for neuromorphic event streams. Similar to FIRES, events are first routed into positive and negative channels to prevent cancellation by an oscillating edge. Next, each channel maintains a per-pixel accumulator using a time-localizing Gaussian filter replicated at periodic spacing to track spectral content close to a target frequency. Here, we define the periodicity of the man-made object:

\begin{equation}
    T = 1/f_{t}
    \label{eq1}
\end{equation}

Where $f_{t}$ is the target frequency for the RTS.

Functionally, this filter is achieved by convolving a Gaussian filter with a comb function at the target frequency in the time domain to enable a user-tunable time-frequency analyzer with bandwidth to compensate for drift in event frequency or any inconsistent frequency characterization. Spectrally, this filter consists of a comb in the frequency domain weighted by a Gaussian envelope. The spectral comb allows spectral content to be supported based on its power at intervals of the target frequency while the Gaussian envelope may be tuned to weight successive harmonics to distinguish orders by their relative weighting. 

To support memory efficient operation, the filter is not required to be explicitly stored in memory and may instead directly calculate its weighting based on the timestamp of an ingested event. For a received event of a given polarity at pixel coordinate $(i,j)$ and at timestamp $t_k$, the RTS applies a per-pixel weighting, $R_{ij}(t_k)$ based on:

\begin{equation}
    R_{ij}(t_k) = exp(-\hat{t_k}^2/(2\sigma^2))
    \label{eq:rts_weighting}
\end{equation}

Where $\sigma$ is the bandwidth of the Gaussian operator and $\hat{t_k}$ is a change of variables designed to map $t_k$ onto its minimum periodic distance to $T$ and takes the form:

\begin{equation}
    \hat{t_k} = |((t_k + T/2) \% T) - T/2|
    \label{eq:timestamp_mapping}
\end{equation}

Where $\%$ is the modulus operator. 

This change of variables weights $t_k$ based on its proximity to the nearest comb spike in the time domain, but assumes that $\sigma << T$ such that neighboring Gaussian-convolved spikes exhibit negligible overlap. For out-of-sync events, the RTS may be modified to apply a static negative value to events outside of the Gaussian envelope (e.g., $\hat{t_k} > 3\sigma$) to penalize events that exhibit undesired frequency content. 

\begin{sloppypar}
\begin{figure}[hbt!]
\centering
\includegraphics[width=.8\textwidth]{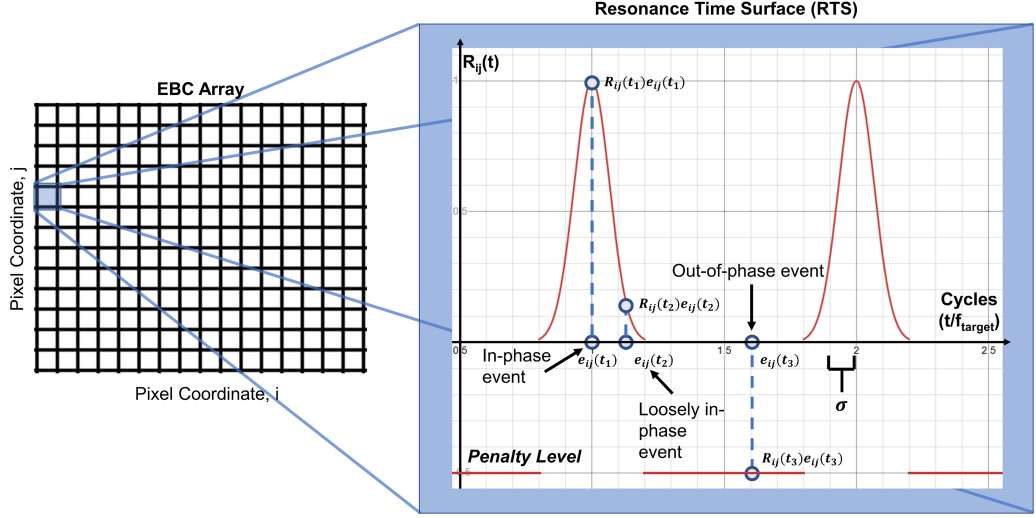}
\caption{Overview of RTS Visualization. Events are ingested pixel-wise and split based on polarity. Ingested events are temporally weighted based on their timestamp and added to the (i,j) pixel location on the RTS based on the event coordinate.}
\label{fig:RTS}
\end{figure}
\end{sloppypar}

As shown in Figure \ref{fig:RTS}, the application of the RTS to an event feed creates three categories of events: in-phase, loosely in-phase, and out-of-phase events. In-phase events are maximally weighted by the RTS to build a surface highly reflective of their contribution. Loosely in-phase events appear within the bandwidth of the Gaussian operator but exhibit lower weighting. Out-of-phase events are either wholly rejected by the RTS or directly penalized to reduce their contribution to the final surface.


\section{Experimental Setup and Data Collection}

\begin{figure}[hbt!]
\centering\includegraphics[width=0.7\linewidth]{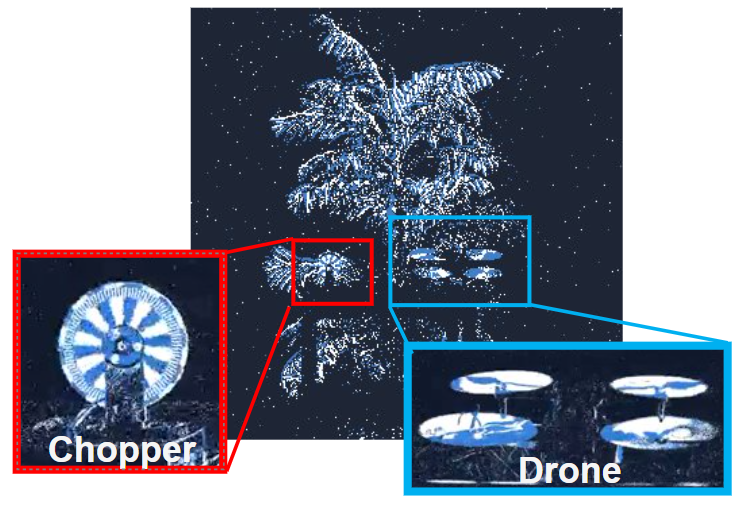}
\caption{Lab experimental set up of the chopper highlighted in red, and a drone mounted to a table top highlighted in blue. A tree sits in the background of the image representing a cluttered environment. }
\label{fig:LabSetUp}
\end{figure}

To test the FRIES pipeline, two data sets are used, collected with a Prophesee EVK4 event-based imaging system. The first data set is from a lab experiment configured to observe a drone rotor and a mechanical chopper inside a controlled environment shown in Figure \ref{fig:LabSetUp}. The chopper is operated at a constant frequency of 110Hz, and the drone is mounted in a fixed position. A dynamic tree is placed behind the drone and chopper to mimic a cluttered environment.  The event-based camera is configured with Bias on = 0 and Bias off = 0, optimal for drone detections based on the vendor's recommendations. All other camera parameters including threshold values, hardware position, and lighting are held constant for the duration of the indoor lab data collection. The drone and chopper are turned on and allowed to stabilize to their operating frequency. Next, the tree is manually agitated to mimic the low-frequency effects of wind and a temporally varying background. The scene is recorded over approximately 10 second intervals.

\begin{sloppypar}
\end{sloppypar}

\begin{figure}[hbt!]
\centering\includegraphics[width=0.7\linewidth]{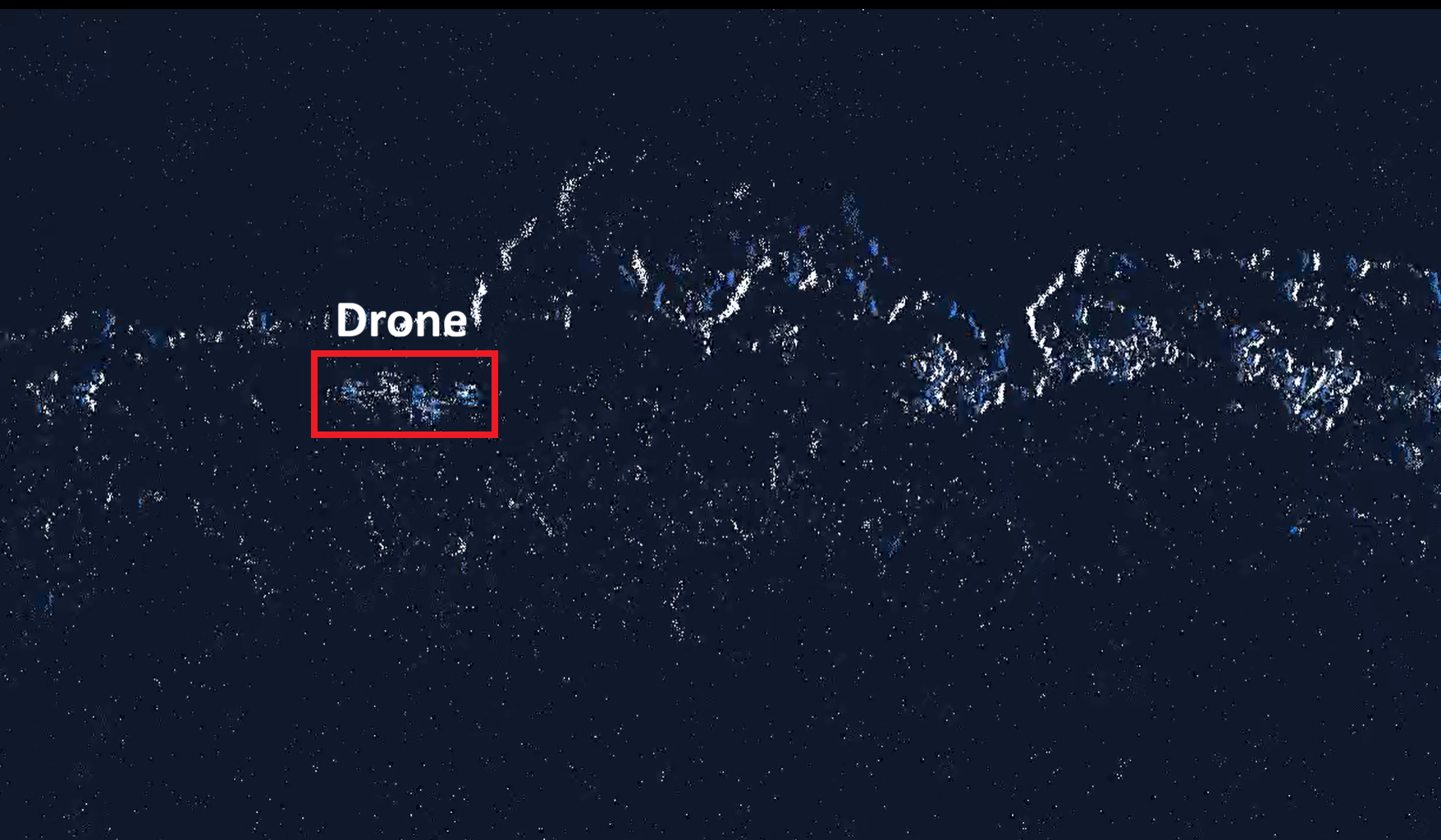}
\caption{Image from the outdoor data collection of a drone, indicated by the red box, hovering in front of a tree line for FRIES pipeline testing in a real world cluttered environment.}
\label{fig:FieldTestSetUp6}
\end{figure}

The second data collection is conducted in an uncontrolled outdoor environment to evaluate FRIES under realistic, dynamic conditions. A class 1 drone is deployed to hover in front of a distant treeline, shown in Figure \ref{fig:FieldTestSetUp6}, creating a challenging scenario in which the target exhibits low visual contrast against a complex, textured background. The drone is operated in a stationary hover at a fixed location and an approximately constant altitude, oriented such that its rotor disk faces the sensors. A nearby treeline provided a cluttered backdrop with natural motion from wind-driven foliage and slowly varying illumination.

\begin{sloppypar}
\end{sloppypar}

\section{Results}
\label{sec:results}

To demonstrate the capability of FRIES to discriminate man-made objects with a consistent frequency or vibration component from a natural background and visualize extracted frequencies using the RTS, the data set from the lab experiment and the outdoor demonstration were used. Each recorded event stream was processed using the FRIES pipeline to determine ROIs and extract the dominant frequency components of each detection candidate. The dominant frequency values were analyzed and the associated drone rotors and chopper detections were manually isolated from false positives driven by noise and background. The corresponding frequencies were subsequently visualized using the RTS to enable selective surveillance of detected targets. By extracting the frequency component from the known chopper rotation and the expected rotor behavior, we can assess the ability of FRIES to discriminate objects containing a frequency signal in cluttered environments.

\subsection{FRIES Results of Lab Experiment}

Using the indoor laboratory experiment data, the temporal gating stage of FRIES appeared to reduce events associated with spurious high-frequency artifacts, as well as slow changes less likely to reflect physically meaningful motion, helping retain events generated from periodic signatures of targets such as rotor edges and the mechanical chopper. The resulting activity maps exhibited a nonuniform spatial distribution of event density where pixels associated with the rotating chopper and drone rotors consistently showed elevated activity relative to the surrounding scene. Using the median activity value as an adaptive threshold helped separate the high-activity target regions of interest from the predominantly lower-activity background across tested conditions, including suppressing background clutter from the dynamic tree. Restricting analysis to these ROI candidates maintained spatial localization and reduced the data passed to the Fourier analysis stage, improving computational efficiency. In these experiments, DBSCAN clustering consistently produced clusters associated with the mechanical chopper and with individual drone rotors, reflecting detection of these surrogate targets in the controlled setting. However, the method’s selectivity and robustness beyond these conditions requires further investigation and tuning.

\begin{figure}[hbt!]
\centering\includegraphics[width=.9\linewidth]{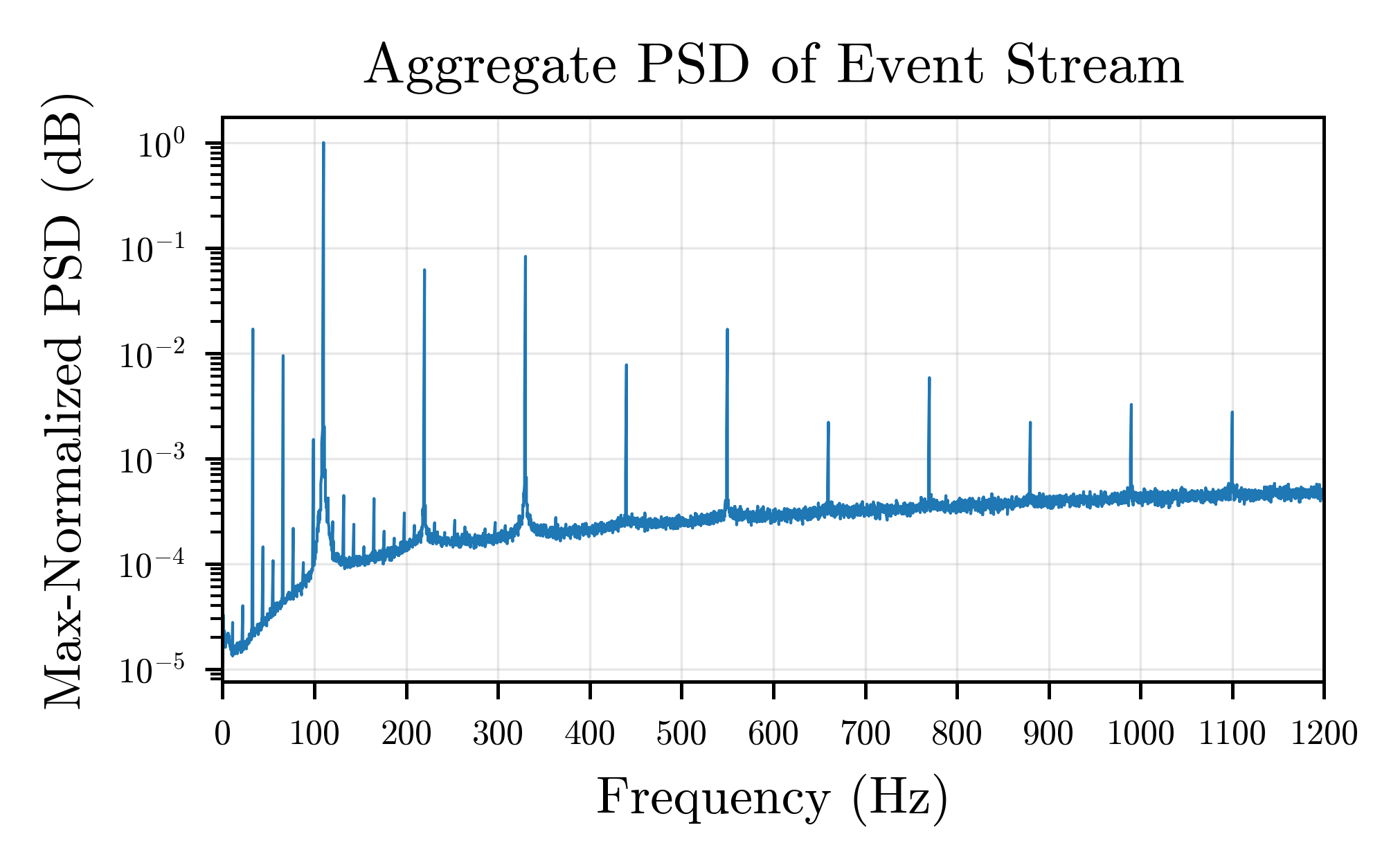}
\caption{Spectral extraction plot using the summed per-pixel fast Fourier transform PSD with a 5 second temporal window.}
\label{fig:frequency_psd_1200_figsize_3.5_2.2}
\end{figure}

\begin{figure}[hbt!]
\centering\includegraphics[width=0.45\linewidth]{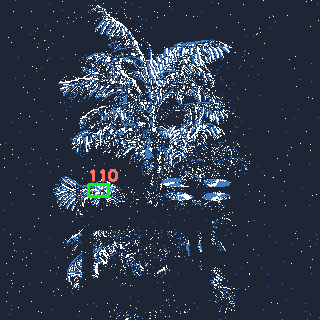}
\caption{Lab experiment scene with a frequency bandpass filter applied to the extracted spectra to visualize exclusively the chopper detection.}
\label{fig:chopper_only_filter_cluster_freq_filter}
\end{figure}

The spectral analysis recovered the mechanical chopper’s rotation frequency of 110Hz, in agreement with the manufacturer’s specified rate and the commanded Revolutions Per Minute in the lab setup. Figure \ref{fig:frequency_psd_1200_figsize_3.5_2.2} shows a plot of the distribution of detected frequencies across all pixels. The Fourier Transform was calculated per-pixel with a 5 second temporal window to compute the PSD for each pixel. The PSDs were then summed across all pixels and the dominant frequencies were plotted. Notably, the plot clearly illustrates the dominant frequency corresponding to the chopper, which appears as a distinct peak in the distribution at 110Hz. For the multi-rotor drone, four distinct rotor ROIs exhibited dominant frequencies between approximately 129–451Hz, while background structures exhibited lower dominant frequencies. Figure \ref{fig:chopper_only_filter_cluster_freq_filter} illustrates the output of FRIES with the implemented frequency-selective filtering, applied to the event stream, detecting only the chopper’s mechanical motion. The validated extracted chopper detections demonstrate FRIES can be used as a tunable spectral gate, enabling targeted visualization of man‑made periodic motion within cluttered event data.

\begin{sloppypar}
\end{sloppypar}

\begin{sloppypar}
\end{sloppypar}

While the drone rotor frequencies could not be independently validated due to the absence of direct measurements, they fall within the expected operating range for similar small multi-rotor platforms \cite{article}, indicating FRIES is extracting physically plausible signatures. In this setup the drone was rigidly mounted rather than freely flying, and the onboard IMU and flight controller continuously attempted to level the vehicle. Because the mount was slightly tilted, the drone could not reach a true level attitude and therefore did not always achieve a stable operating point. As a result, one rotor ROI intermittently dropped out over the course of the recording. This behavior is consistent with the FRIES assumption of quasi-stationary periodicity; when a rotor undergoes rapid or irregular speed changes, its dominant frequency becomes nonstationary over the batch duration, causing its spectral peak to decorrelate and the ROI to be rejected as a nonpersistent or false detection.

\subsubsection{RTS Visualization}
\label{sec:rts_vizualization}
\begin{figure} [hbt!]
\centering\includegraphics[width=\linewidth]{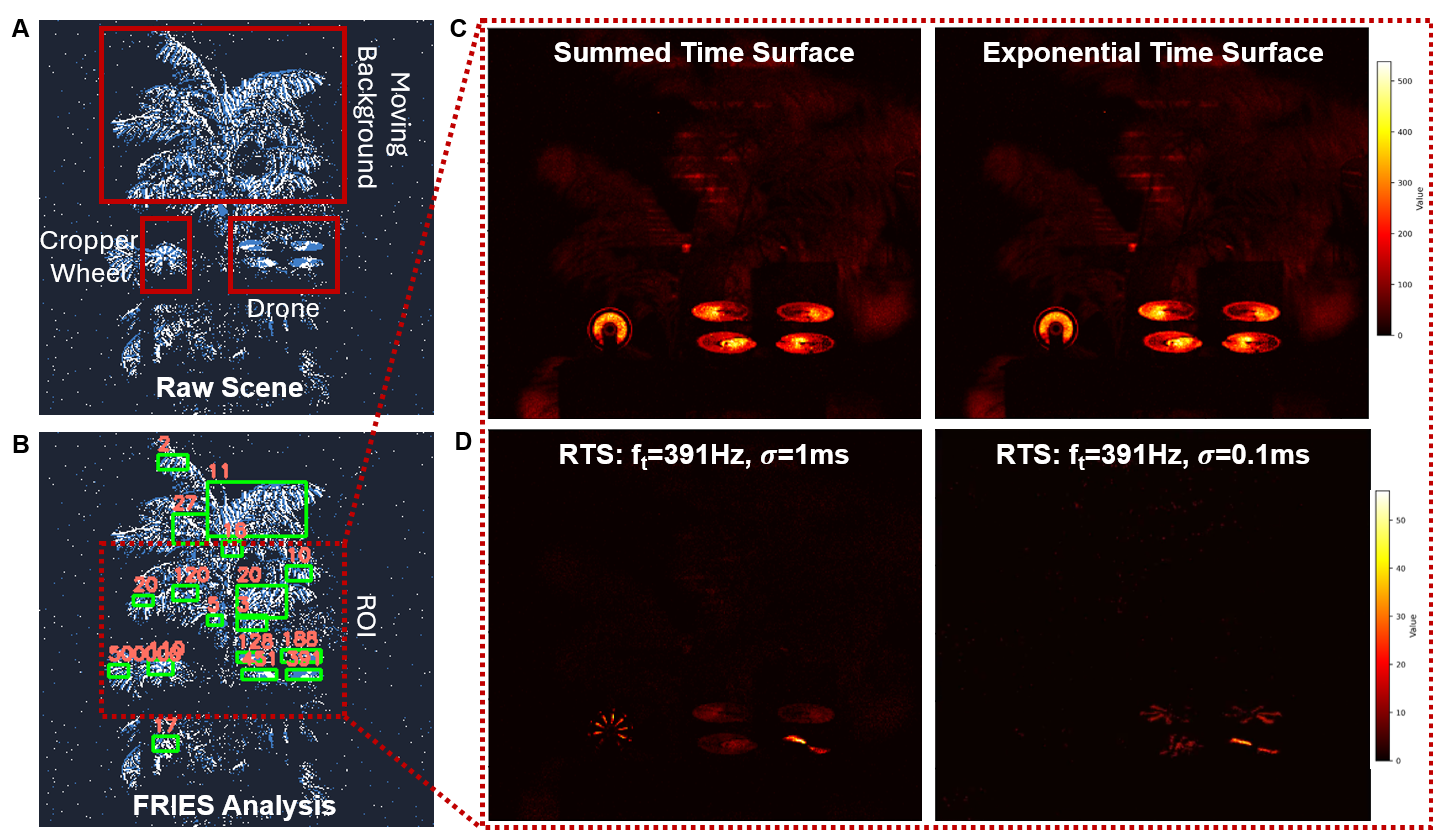}
\caption{Detection of motorized targets in tabletop experiment. (A) Description of experimental setup and targets. (B) FRIES extracted frequency signatures. (C) RTS of the drone and chopper at a target frequency of 391Hz and a Gaussian width of 1ms and (D) 0.1ms, respectively, for enhanced selectivity.}
\label{RTS}
\end{figure}

After processing, the positive events within a subset of the scene were reshaped into time surfaces to emphasize features captured by each visualization method. Negative events were rejected as they exhibited a higher noise floor, leading to time surfaces with less feature relevance than their positive counterparts. The ROI was selected to contain the chopper wheel, drone, and a portion of the background to offer a subset of the event stream with diverse frequency content. As shown in Figure \ref{RTS}C, the ROI was shaped into a summed time surface, exponential time surface, and two RTSs, where the RTSs were tuned to extract the bottom right drone rotor at a measured frequency of 391Hz with a Gaussian envelope containing a standard deviation of 1ms and 0.1ms, respectively, to enable more broad and selective frequency extraction. The RTS-extracted frequency of 391Hz matched the FRIES frequency estimate for the corresponding drone rotor shown in Figure \ref{RTS}B. Due to the lack of frequency weighting, the summed time surface and exponential time surface blur the motorized targets due to their dense spatio-temporal event signature. In contrast, the frequency-sensitive nature of the RTS enables extraction of the selected drone rotor while suppressing unstructured background events and non-resonant features from other targets, confirming its effectiveness as a selective surveillance tool.

\begin{sloppypar}
\end{sloppypar}

\subsection{FRIES Results of Outdoor Demonstration}
\label{sec:field test}

\begin{figure} [hbt!]
\centering\includegraphics[width=\linewidth]{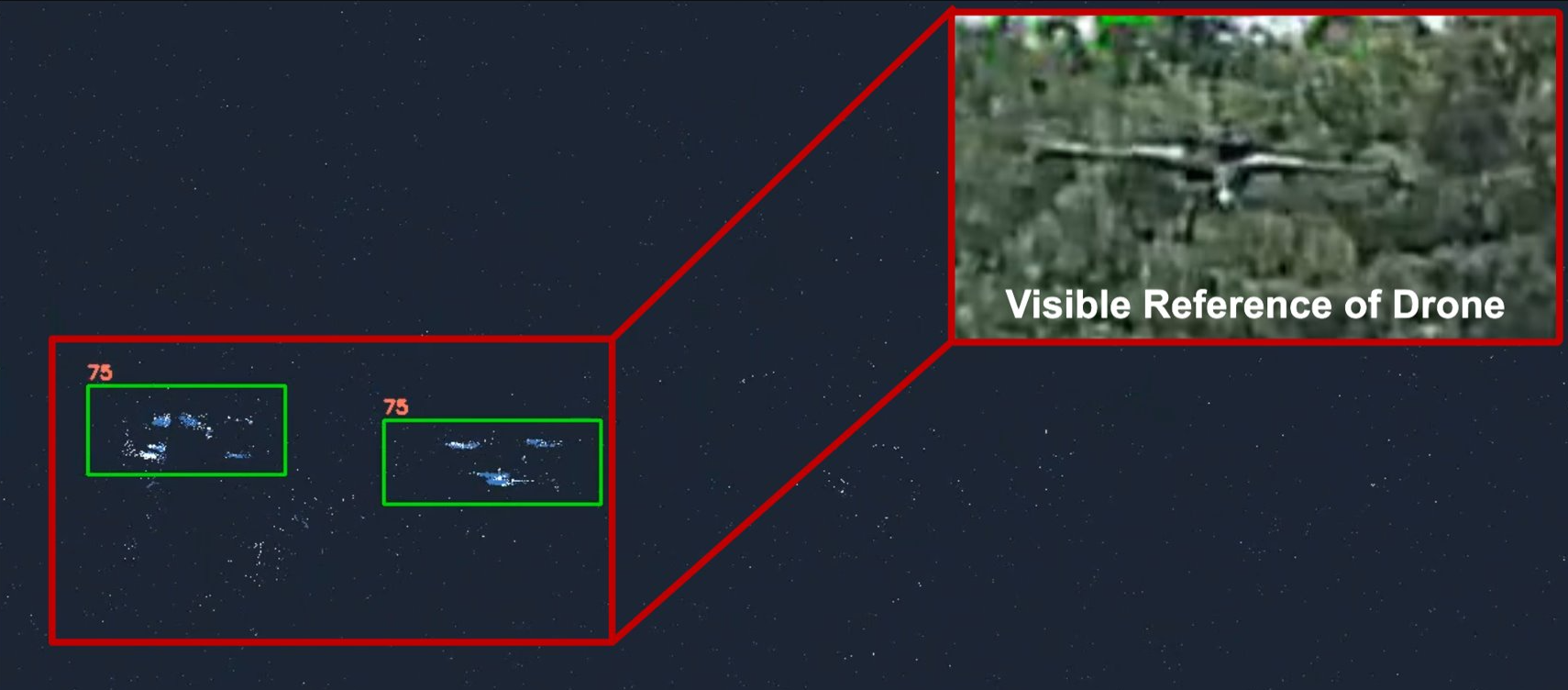}
\caption{Overlaid frequency values found for the cropped outdoor tested drone using the FRIES algorithm including a visible snapshot of the drone using a frame-based camera at the same time frame for reference purposes.   }
\label{fig:DroneFieldTest}
\end{figure}

\begin{figure}[hbt!]
\centering\includegraphics[width=.9\linewidth]{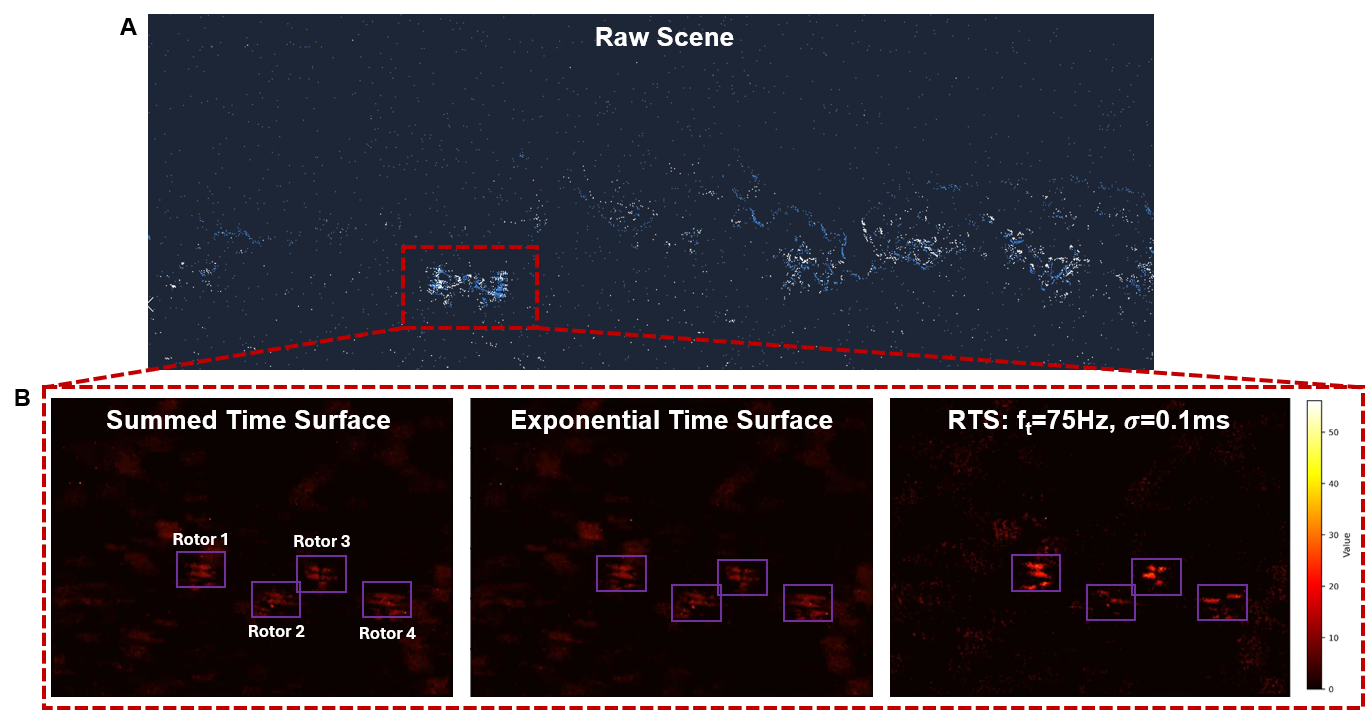}
\caption{Discrimination of industrial drone during outside collection. (A) Panoramic of a drone in flight against a treeline. (B) Time surface representations of the drone against the treeline with RTS demonstrating greater selectivity of drone events.}
\label{fig:rts_overview_viz}
\end{figure}

The second experiment demonstrates FRIES in an uncontrolled outdoor environment. A drone was deployed to hover in front of a treeline; imaged simultaneously by a conventional frame-based sensor and an event-based sensor shown in Figure \ref{fig:DroneFieldTest}. Under this background, uncontrolled low-frequency, spatially distributed activity generated event-camera responses comparable to the full number of events to those produced by the drone, as shown through the summed time surface magnitude in Figure  \ref{fig:rts_overview_viz}, thereby testing FRIES’s ability to separate man-made, periodic motion from natural, unstructured clutter under high signal-to-background conditions.

\begin{sloppypar}
\end{sloppypar}

\begin{figure}[hbt!]
\centering\includegraphics[width=\linewidth]{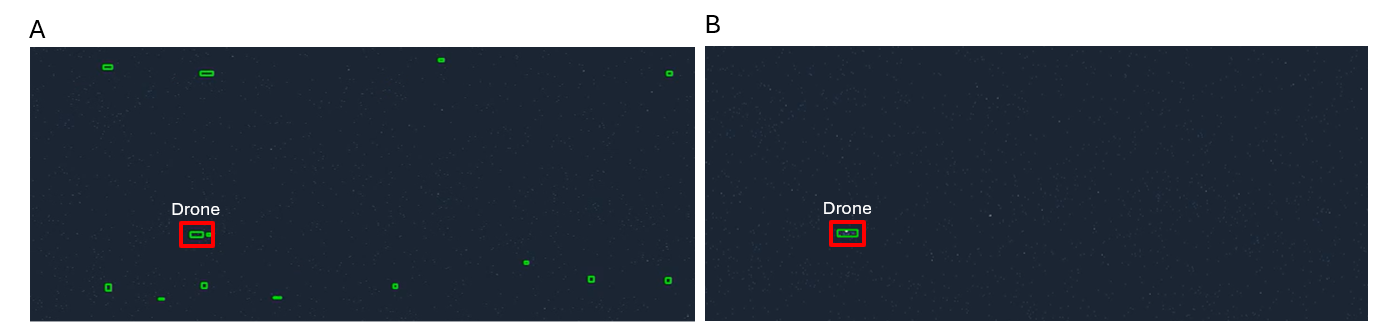}
\caption{Outdoor data collect of the drone with and without the FRIES time gating step. Image A shows the scene without any time gating, including false positive candidate detections. Image B shows the scene after the time gating where noise has been suppressed.}
\label{fig:TimeGate}
\end{figure}

From inspection, the event-frame with FRIES processing extracts the drone rotor from the background with high contrast, when compared to the visible-frame image where the drone exhibits low contrast against the complex treeline. This result emphasizes the event-based data processed with FRIES reveals the drone rotors by leveraging their periodic signature as a detection prior rather than relying on spatial contrast alone. The same time gating parameters used in the laboratory experiments were applied without modification. 
Through the time gating operation, illustrated in Figure \ref{fig:TimeGate}, we observe a clear reduction in spurious detections within the outdoor scene. In Image A, taken without the FRIES time gating step, multiple false positive candidate detections appear alongside the drone, likely caused by noise or background motion. In contrast, Image B displays the same scene following the application of time gating, where non-drone events have been suppressed. As a result, the drone detection is more prominent and isolated, demonstrating the utility of temporal gating for reducing false positives in complex, dynamic outdoor environments. From these gated events, the median-based threshold and DBSCAN clustering again yielded a small number of compact ROIs. In Figure \ref{fig:DroneFieldTest} the dominant cluster was spatially aligned with the drone body and rotor disk, while diffuse background activity either failed to meet the density criterion and was rejected as noise or formed small, isolated clusters that were pruned in later spectral stages. 

\begin{sloppypar}
\end{sloppypar}

Applying local spectral analysis to the principal ROI produced a dominant frequency near 75Hz, interpreted as the rotor rotation frequency of the hovering drone. Concurrently, background ROIs associated with tree motion exhibited lower dominant frequencies in the 2–27Hz range, corresponding to slow foliage undulations driven by wind. This separation in frequency content enabled FRIES to discriminate the drone’s relatively narrowband, higher-frequency oscillatory signature from the broad, low-frequency background in this image. The 75Hz estimate is consistent with expected rotor rates for a hovering multi-rotor platform \cite{article}, although it could not be independently validated due to the absence of ground-truth instrumentation and required parameter tuning to achieve consistent extraction with reduced false positives.

\begin{figure}[htbp]
  \centering
  \begin{subfigure}[b]{0.32\textwidth}
    \includegraphics[width=\linewidth]{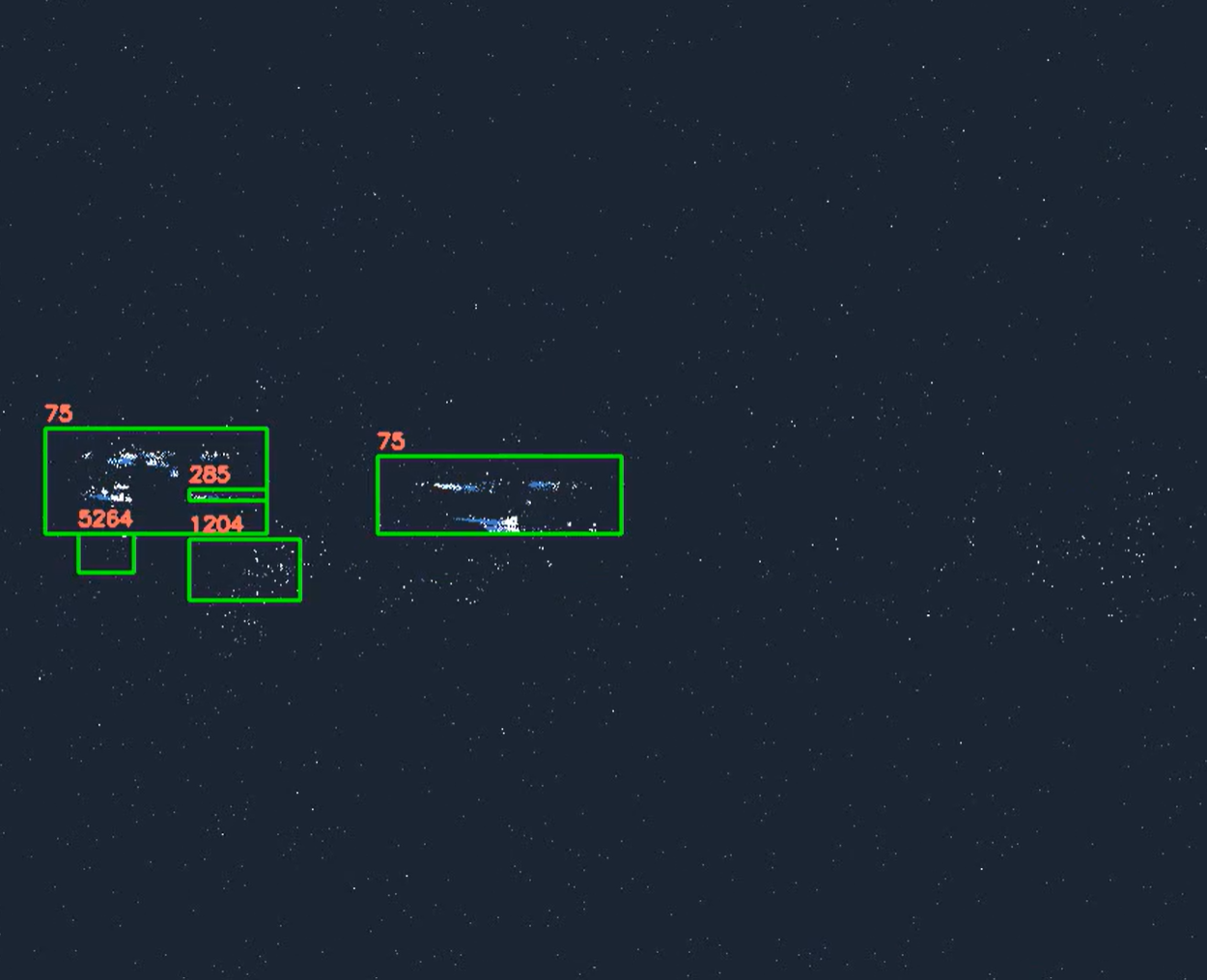}
    \label{fig:fieldbad4}
  \end{subfigure}
  \hfill
  \begin{subfigure}[b]{0.32\textwidth}
    \includegraphics[width=\linewidth]{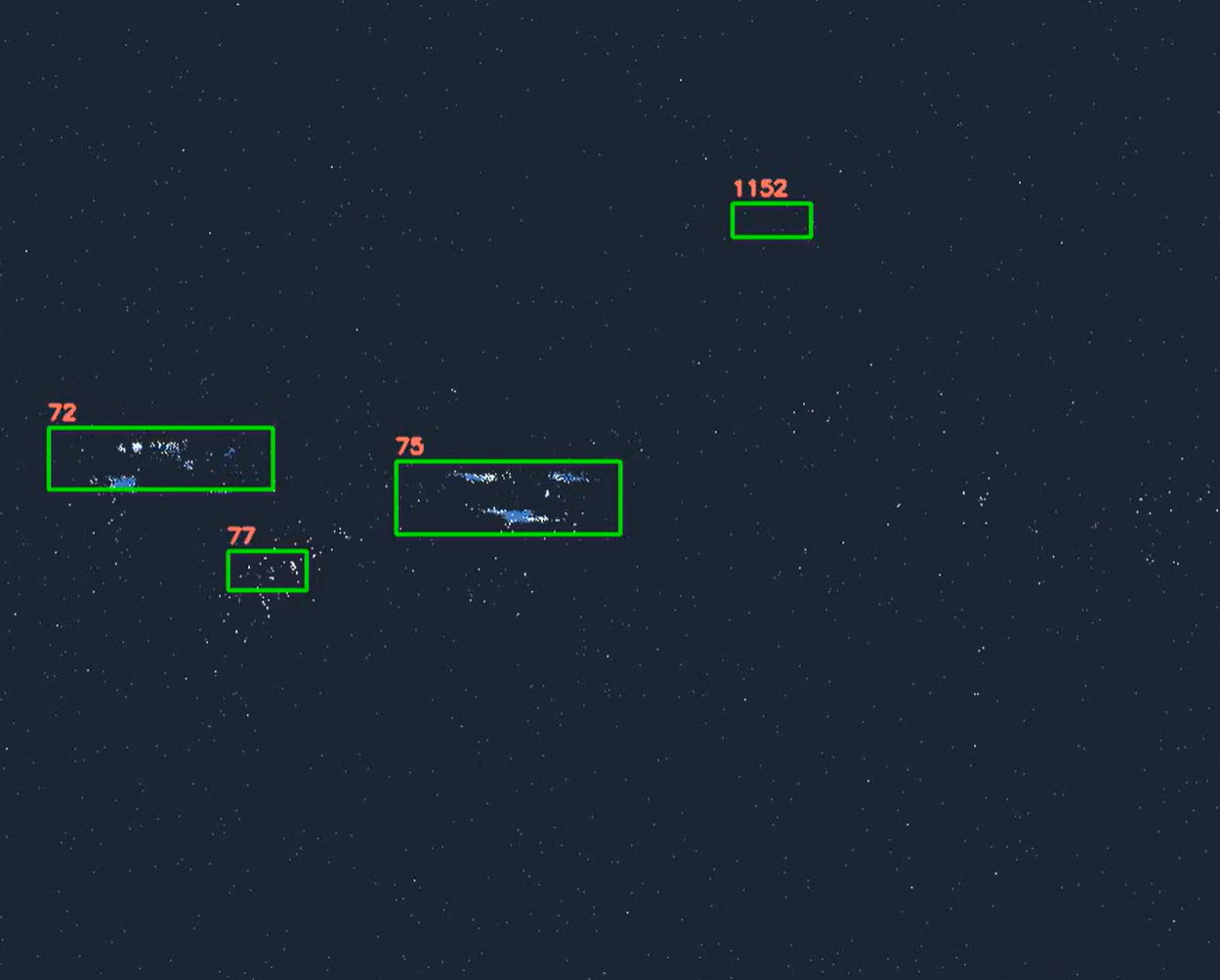}
    \label{fig:fieldbad2}
  \end{subfigure}
  \hfill
  \begin{subfigure}[b]{0.32\textwidth}
    \includegraphics[width=\linewidth]{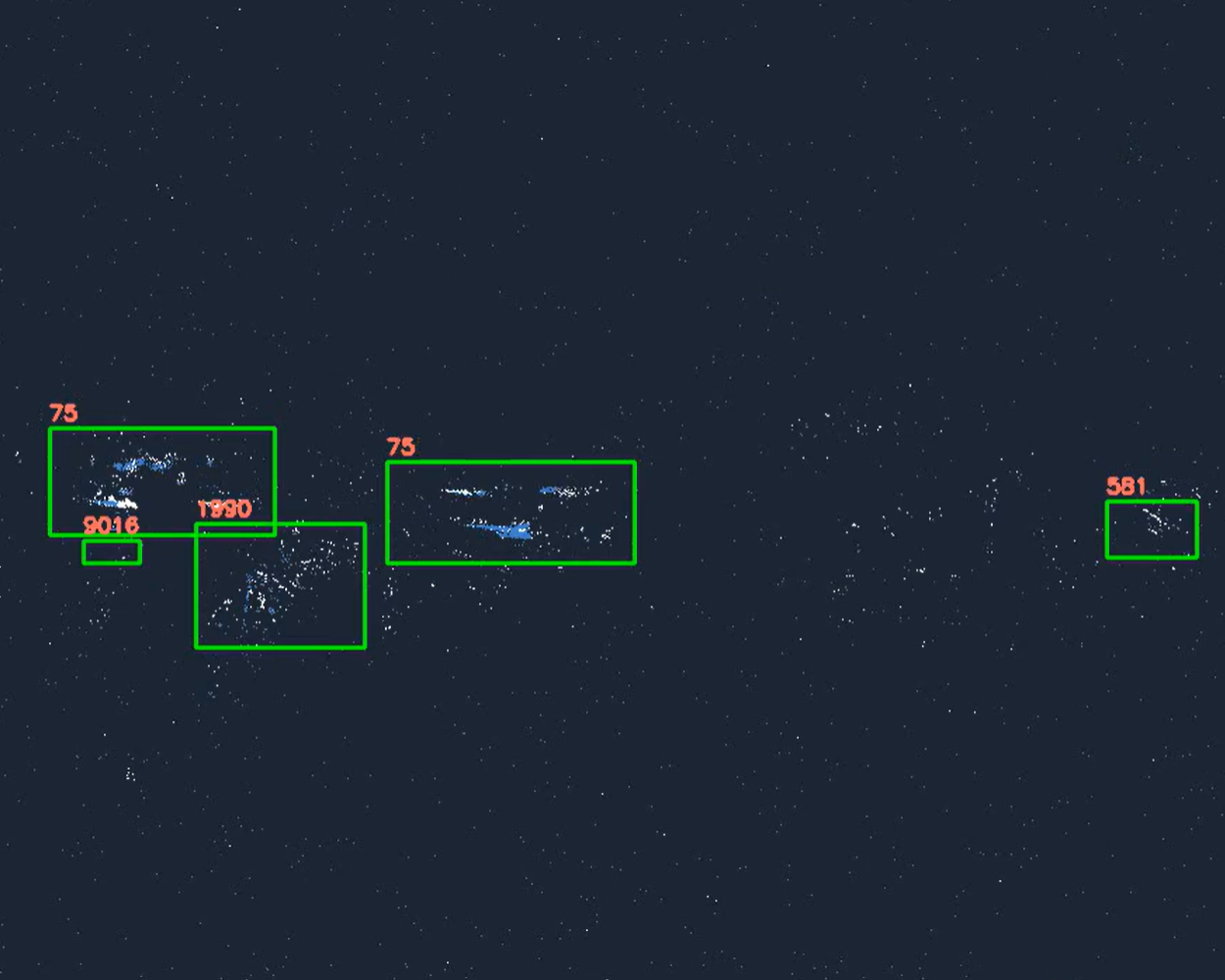}
    \label{fig:fieldbad3}
  \end{subfigure}
  \caption{Three distinct timepoints from the FRIES output on the outdoor dataset reveal additional detections beyond the filtered drone region, indicating the algorithm requires further tuning to reduce false positives.}
  \label{fig:three_images}
\end{figure}

The outdoor experiment presented in Figure \ref{fig:three_images} illustrates both the promise and current limitations of the FRIES algorithm. Three selected FRIES output frames reveals imperfect selectivity. In addition to the consistent detection of the drone rotors, the algorithm intermittently identified false positive regions. These spurious detections, corresponding to background clutter, noise or momentary motion in the foliage, were not persistent and appeared sporadically throughout the video. These false positives demonstrate that, while FRIES is capable of extracting chopper and drone oscillatory signatures, its performance in uncontrolled, outdoor settings is not yet reliable for operational use. The laboratory experiment produced consistent extractions, while the outdoor field test resulted in a mix of true detections and sporadic false positives. The outdoor results are ultimately inconclusive and highlight the necessity for further tuning, potentially including adaptive thresholding, environmental calibration, or enhanced spectral discrimination to improve robustness across diverse scenes.

After detection, the candidate ROI was selected and processed by the summed time surface, exponential time surface, and RTS with a target frequency of 75Hz and Gaussian envelope with a standard deviation of 0.6ms, as shown in Figure \ref{fig:rts_overview_viz}B. In the summed and exponential time surface, the events generated by the drone exhibit a comparable density as those generated by the background, leading to poor identification of the drone. In contrast, the RTS suppressed the unstructured background frequency events, leading to an approximate 4x improved ratio between background RTS values and rotor RTS values to provide enhance distinction from the background. This result reinforces the utility of the RTS to extract key features of motorized objects in natural scenes. These experiments validate the fundamental approach of FRIES and RTS for frequency-based detection of mechanical targets, but also illustrate the challenges of applying the algorithm in dynamic, cluttered environments.

\begin{sloppypar}
\end{sloppypar}

\section{Discussion}
FRIES provides a frequency-domain analysis framework for event-based data to detect objects exhibiting strong spectral signatures, such as those arising from man-made objects, from a natural background. In tabletop experiments, FRIES recovered a chopper wheel at its manufacturer-specified frequency of 110Hz and detected distinct frequencies to each drone rotor between 129–451Hz, while appropriately attributing lower-frequency activity to a dynamic tree. While the drone rotors exhibited inconsistent frequencies, this artifact arises from the drone being clamped to the table, triggering the stabilization routine to drive the rotors at different frequencies. When applied to outdoor trials, FRIES was able to extract two drone rotors against a wind-driven treeline at a simultaneous 75Hz, but requires parameter tuning and algorithm improvements for consistent detections in uncontrolled environments. After extraction, individual frequency content was enhanced through the RTS to selectively surveil the specific mechanical sources while suppressing background and out-of-sync content. The EBC detection algorithm, FRIES, offers a potential advantage in detecting objects with faint contrast and enables spectral analysis based detection strategies that would otherwise require very high frame rate conventional cameras. However, EBCs trade salient features to enable this imaging modality \cite{greene_pytorch-enabled_2025}, reinforcing the perspective that EBCs may prove valuable as a complementary system to conventional frame-based cameras rather than an alternative \cite{gehrig_low-latency_2024}. 

The key innovation of FRIES is the introduction of a frequency-driven mechanism as a primitive for detection versus reliance on purely spatial or temporal trends. By time gating and using an activity map as a prior for detection, FRIES suppresses noise and reduces the demand to conduct full field spectral analysis to selected ROIs, which leverages the assumption that high-frequency objects will produce a greater density of events than the surrounding low-frequency background. By calibrating to a median background activity rate, FRIES operates effectively in scenarios where targets are sparse against the background, a likely scenario in wide-field-of-view surveillance tasks, and purely requires the EBC parameters to encode some signal from the target, versus requiring exact or optimized parameters. This approach functions by treating each event timestamp sequence as a byproduct of the underlying physical behavior and rewarding periodicity as an indicator for man-made activity. The RTS extends this functionality by providing a visualization tool to filter and structure incoming events based on their phase coherence with an identified frequency, leveraging discovered spectral knowledge as a prior for selective surveillance and lowering computational needs. 

In its current design, several choices in FRIES are heuristic and indicate future advancements on the framework. The temporal gating parameters were selected empirically based on observed rotor frequencies and were not derived automatically from the discovered frequencies. Additionally, the use of a bank of bandpass filters spanning narrow frequency ranges could help prioritize spectral content ranges to analyze and enable the use of band-limited spectral analysis \cite{bonami_band-limited_1993}. Next, the median-based activity threshold serves as a coarse estimator of background activity and may degrade on dense or textured scenes \cite{magrini_neuromorphic_2025}, necessitating replacements with running estimates over longer temporal windows. Finally, the use of DBSCAN and the fast Fourier transform (FFT) operator operates on batches of events, creating latency on the otherwise incremental event stream. Replacing DBSCAN with an incremental spatial clustering method and replacing the FFT with a recursive spectral operator, such as a sliding window discrete Fourier transform \cite{jacobsen_sliding_2003}, would reduce detection latency while preserving the frequency-based paradigm. While individual operators may be exchanged, we emphasize the objective is to maximize the efficiency of the four-stage (e.g., filter, cluster, spectral analysis, visualize) FRIES pipeline that underpins the spectral-driven detection scheme.

The most immediate extension of this project is to address the remaining gaps between the current prototype and a deployable system. First, FRIES is currently not compared to existing event-based detection algorithms or any frame-based camera algorithm, and establishing or leveraging open source datasets \cite{magrini_fred_2025} to baseline performance will provide validation as well as an opportunity for algorithmic fusion between two sensing modalities for enhanced capabilities. Additionally, sensor contrast threshold (i.e., bias) dictates the events generated across a target and background \cite{greene_pytorch-enabled_2025}, controlling the discernibility of a rotor signature. Enabling a closed-loop system that uses the spectral strength of FRIES as feedback to adapt the bias settings would stabilize detection performance across ranges and illumination conditions without manual intervention.

Building on these improvements, the next phase will focus on the comprehensive development of the FRIES pipeline to enable real-time operation, as well as advanced tracking and classification capabilities. Key priorities include enhancing the detection performance of the FRIES algorithm for both real-time and real-world applications, and addressing computational challenges such as implementing rolling methods to reduce the substantial processing demands of the Fourier transform calculation. Additionally, broadening the scope of FRIES to function effectively in dynamic, moving environments (rather than only static scenes) will require adapting the spectral analysis detection algorithm for moving platforms, expanding the system’s modality, and determining optimal settings to suppress events generated by motion in the imagery. This work will also involve assembling and labeling an extensive database of target objects for classification purposes.

Finally, the RTS is currently tuned to single-frequency targets, limiting simultaneous multi-target surveillance. Extending to a bank of RTS-inspired filters would support parallel tracking of multiple targets as well as track frequency drift independent of changing rotor speed or anomalous mechanical behavior. As a whole, these extensions would advance FRIES from a proof-of-concept frequency analysis tool towards a low-latency, self-calibrating surveillance paradigm for operation in deployable uncontrolled environments. 

\appendix    

\acknowledgments 
 
We would like to thank Georgia Tech Research Institute's (GTRI) Independent Research and Development (IRAD) funds for supporting this work.  

\bibliography{Joes_Citations, Joes_Citations_2} 
\bibliographystyle{spiebib} 

\end{document}